\documentstyle[12pt]{article}

\def \double  {\baselineskip=  2  \normalbaselineskip}
\newcommand{\beq}{\begin{equation}}
\newcommand{\eeq}{\end{equation}}
\newcommand{\beqa}{\begin{eqnarray}}
\newcommand{\eeqa}{\end{eqnarray}}
\newcommand{\beqan}{\begin{eqnarray*}}
\newcommand{\eeqan}{\end{eqnarray*}}
\newcommand{\no}{\nonumber}

\newcommand{\ul}{\underline}
\newcommand{\ol}{\overline}
\newcommand{\ra}{\rightarrow}

\newcommand{\ben}{\begin{enumerate}}
\newcommand{\een}{\end{enumerate}}
\newcommand{\bfl}{\begin{flushleft}}
\newcommand{\efl}{\end{flushleft}}
\newcommand{\ba}{\begin{array}}
\newcommand{\ea}{\end{array}}
\newcommand{\btab}{\begin{tabular}}
\newcommand{\etab}{\end{tabular}}
\newcommand{\bit}{\begin{itemize}}
\newcommand{\eit}{\end{itemize}}

\newcommand{\cA}{{\cal A}}

\newcommand{\vs}{\vspace}
\newcommand{\hs}{\hspace}
\newcommand{\ld}{\lambda}

\def \sam {\triangle}
\newcommand{\prepr}[1] {\begin{flushright} {\bf #1} \end{flushright} \vskip
1.5cm}
\newcommand{\titul}[1] {\begin{center}{\Large {\bf #1 } } \end{center}\vskip
1.cm}

\newcommand{\autor}[1] {\begin{center} {\bf \lineskip .3cm #1  }
                        \end{center} }

\newcommand{\lugar}[1] {\begin{center}  {\large \it #1   } \end{center}}
\newcommand{\abstr}[1] {{\begin{center} \vskip .5cm {\bf Abstract
                        \vspace{0pt}} \end{center}}\begin{quote} #1
                        \end{quote}}
\double
\newcounter{muni}

\begin{document}
\vspace{.4cm}
\hbadness=10000
\pagenumbering{arabic}
\begin{titlepage}
\prepr{Preprint hep-ph/940XXXX\\PAR/LPTHE/94-32  \\20 August 1994 }
\titul{ Testing factorization in colour-suppressed beauty decays
with the ${\bf B \ra \eta_c + K \hs{2mm} ( K^* ) }$ modes}
\autor{ M. Gourdin\footnote{\rm Postal address: LPTHE, Tour 16, $1^{er}$ Etage,
Universit\'e Pierre {\it \&} Marie Curie and Universit\'e Denis Diderot, 4
Place Jussieu, F-75252 Paris CEDEX 05, France.},
Y. Y. Keum\footnote{\rm Postal address: LPTHE, Tour 24, $5^{\grave{e}me}$
Etage, Universit\'e Pierre {\it \&} Marie Curie and Universit\'e Denis Diderot,
2 Place Jussieu, F-75251 Paris CEDEX 05, France.}
and X. Y. Pham$^*$ }

\lugar{Laboratoire de Physique Th\'eorique et Hautes Energies,\footnote{\em
Unit\'e associ\'ee au CNRS URA 280}
 Paris, France }

\begin{center}
{\small  E-mail : gourdin@lpthe.jussieu.fr, keum@lpthe.jussieu.fr and
pham@lpthe.jussieu.fr, }
\end{center}

\vs{-12cm}
\thispagestyle{empty}
\vs{120mm}
\noindent
\abstr{
The large discrepancy recently observed between $B \ra J/\Psi + K (K^*)$ data
and  commonly used hadronic form factor models might be due to a breakdown of
the factorization assumption.
By showing that the $B \ra \eta_c + K (K^*)$ rates can be calculated without
relying on any form factor model, we propose a test of the factorization
hypothesis.
This test is free of possible uncertainties caused by final state interactions
and $W$-exchange graph contributions.
If factorization holds, the measurable ratio
$\Gamma(B \ra \eta_c + K^*)/\Gamma(B \ra \eta_c + K )$
 is predicted to be in the range  from 0.18 to 0.95 when the CLEO II data on
$B \ra J/\Psi + K (K^*)$ are used.
}
%

\vs{50mm}
{\bf PACS index 13. 25. Hw, 14. 40. Nd }
\end{titlepage}

\newpage
\hspace{1cm} \large{} {\bf I \hs{3mm} Introduction}    \vspace{0.5cm}
\normalsize

The study of $B$ meson decays into charmonium states offer a nice opportunity
for a better understanding of three topics :

\vs{2mm} \hs{10mm} 1. the colour-suppressed mechanism in the beauty sector and
a possible breakdown of the factorization hypothesis \cite{R1} ;

\vs{2mm} \hs{10mm} 2. the hadronic form factors ( associated , at the basic
level, with the heavy  to light quark transitions, namely $b \ra s$ ) that
enter in many rare decay modes of the $B$ mesons, these modes being generally
considered as a nice way to investigate new physics beyond the standard
electroweak model ;

\vs{2mm} \hs{10mm} 3. the violation of CP manifested through one of the three
CP violating angles - usually called $\beta$ - that can be measured by
comparing $B^o$ and $\ol{B}^o$ decays.

\vs{2mm}
The charmonium states experimentally observed to date in both $B^o_d$ and
$B^+_u$ decays are $J/\Psi$, $\Psi'(3685)$ and $\chi_{c_1}(3510)$ accompanied
mostly with $K$ and $K^*(892)$ mesons \cite{R2}$-$\cite{R5}.
The $B \ra \eta_c + X$  inclusive decay as well as  some exclusive modes are
under  experimental investigation \cite{R6}. When available, both the spin 0
and the spin 1 parts of the weak ${\sam}B$ = $- {\sam}S$ = 1 vector and axial
vector currents will be studied.
The measurement of $B \ra \eta_c + K (K^*)$ rates is also a powerful way to
extract the pseudoscalar decay constant $f_{\eta_c}$ \cite{R7}.
Furthermore, as  will be explained in this paper, by combining $B \ra \eta_c +
K (K^*)$ and $B \ra J/\Psi + K (K^*)$ data, the factorization assumption in
colour-suppressed  beauty decays can be tested.

\vs{3mm}
Indeed it has been shown recently \cite{{R1},{R8}}, that assuming
factorization, the decay modes $B \ra J/\Psi + K $ and $B \ra J/\Psi + K^*$
when taken together appear to be two remarkable filters to eliminate many - if
not all - of the commonly used models for $B \ra K (K^*)$ hadronic form factors
as given by quark models \cite{R9}$-$\cite{R15}, by QCD sum rules \cite{R16},
by lattice calculations \cite{R17}, by  union of chiral and heavy flavour
symmetries \cite{R18} and also those derived in Refs.\cite{{R1},{R8}} employing
the Isgur-Wise procedure \cite{{R19},{R20}} which relates the $B \ra K (K^*)$
form factors to the experimentally measured $D \ra K (K^*)$ ones, using the
heavy flavour symmetry between  $b$ and $c$ quarks.

It is interesting to observe that the only experimental data necessary for
these exclusions are the relative ones.

\vs{2mm}
\hs{10mm} 1) the ratio of $K^*$ versus $K$ production
\beq
R_{\Psi} = \frac{\Gamma(B \ra J/\Psi + K^*)}{\Gamma(B \ra J/\Psi + K)} \
\label{eq1}
\eeq

\vs{2mm}
\hs{10mm} 2) the fractional longitudinal polarization in the $J/\Psi K^*$ mode
\beq
\rho_L = \frac{\Gamma(B \ra J/\Psi + K^*)_{LL}}{\Gamma(B \ra J/\Psi + K^*)} \
\label{eq2}
\eeq

\newpage
The experimental data are available for both $R_{\Psi}$ and $\rho_L$
\beqa
R_{\Psi} &=& 1.64 \pm 0.34 \hs{22mm}
{\rm ARGUS, CLEO \hs{1mm} 1.5, \hs{1mm} CLEO \hs{1mm} II \hs{2mm} Averaged
\hs{3mm} \cite{R2}}  \no \\
R_{\Psi} &=& 1.71 \pm 0.34 \hs{22mm} {\rm CLEO \hs{1mm} II \hs{5mm} \cite{R3}}
\no \\
\cr
\rho_L &>& 0.78 \hs{5mm}(95\% \hs{2mm} CL) \hs{11.5mm}
{\rm ARGUS \hs{5mm} \cite{R4}} \no \\
\rho_L &=& 0.80 \pm 0.08 \pm 0.05 \hs{10mm}
{\rm CLEO \hs{1mm} II \hs{4.5mm} \cite{R3}}  \no \\
\rho_L &=& 0.66 \pm 0.10 \hs{2mm}^{+ 0.10}_{- 0.08} \hs{13mm}
{\rm CDF \hs{10.5mm} \cite{R5}}  \no
\eeqa

The most spectacular feature of these data is the large value of $\rho_L$ and
the  main difficulty is to account simultaneously for both $R_{\Psi}$ and
$\rho_L$.
For instance the modified Bauer-Stech-Wirbel model discussed in \cite{R1} as
BSW II (denoted as "revised" BSW in Ref.\cite{R3} ) produces correctly
$R_{\Psi} = 1.61$ and uncorrectly $\rho_L = 0.35$, i.e. a value too small by a
factor of \hs{1mm} 2.
On the other side, the so called "alternative model"  of Ref.\cite{{R7},{R15}}
constructed so that to have a large value for $\rho_L$, fails badly to account
for $R_{\Psi}$,
predicting a value 5 times larger than the experimental one.

\vs{3mm}
When we realized \cite{R1} the large disagreement between experiments
\cite{R2}$-$\cite{R5} and model predictions \cite{R9}$-$\cite{R18}, including
those derived in \cite{{R1},{R8}} from the Isgur-Wise scaling method
\cite{R19}, we were naturally led to ask the question what might be the origin
of such a discrepancy ?

We proposed two possibilities :

\vs{2mm} \hs{5mm} 1) an inadequacy of the ingredients - normalization at $q^2
=0$, dependence with respect to $q^2$ - used in the $B \ra K (K^*)$ hadronic
form factors ;

\vs{2mm} \hs{5mm} 2) a failure of the factorization assumption.

\vs{2mm}
The purpose of this paper is to investigate the second possibility by proposing
a test of factorization.
We show that the $B \ra \eta_c + K (K^*)$ and $B \ra J/\Psi + K (K^*)$ rates
can be related assuming only factorization and plausible scenarios for the
$q^2$ dependence of some ratios of form factors.
The computation does not rely on any form factor model.
Measurement of the $\eta_cK$ and $\eta_cK^*$ decay modes of $B$ mesons will
provide a genuine test of factorization if data are accurate enough.

\vs{7mm}
\large{} {\bf II \hs{3mm} Computation of the $B \ra \eta_c(J/\Psi) + K(K^*) $
decay widths}
\vspace{0.5cm}
\normalsize

1$^{\bullet}$) \hs{3mm} To start with, let us remind the $QCD$-induced
effective Lagrangian \cite{R21} relevant for the processes considered here.
\beq
{\cal L}_{eff} = \frac{G_F}{\sqrt{2}} \ V_{cb}^* V_{cs}
\{ c_1 [\ol{b} \Gamma^{\rho} c][\ol{c} \Gamma_{\rho} s] +
c_2 [\ol{b} \Gamma^{\rho} s][\ol{c} \Gamma_{\rho} c] \} + H. C. \label{eq3}
\eeq
where $\Gamma_{\rho} \equiv \gamma_{\rho}(1 - \gamma_5)$ corresponds to V-A
currents.
Using a Fierz transformation, the Lagrangian (3) can be rewritten for
colour-suppressed processes as
\beq
{\cal L}_{eff} = \frac{G_F}{\sqrt{2}} \ V_{cb}^* V_{cs}
\{ (c_2 + \frac{1}{N_c} \ c_1) [\ol{b} \Gamma^{\rho} s][\ol{c} \Gamma_{\rho} c]
+
c_1 [\ol{b} \Gamma^{\rho} T^a s][\ol{c} \Gamma_{\rho} T^a c] \} + H. C.
\label{eq4}
\eeq
where the $T^a$'s are SU(3) colour generators.

The brackets in the first term of ${\cal L}_{eff}$ are colour singlet currents
while in the second term they are colour octet currents.
In the factorization approximation of Bauer, Stech and Wirbel \cite{R9}
(henceforth BSW), physical hadronic states are inserted in all possible ways.
These states being colourless, only the first part of the Lagrangian (4)
survives and the quantity $a_2 = c_2 + \frac{1}{N_c} \ c_1 $ \hs{1mm} is
considered as a purely phenomenological parameter.
The second part of the Lagrangian (4), sometimes called non-factorizable
contribution, might happen to cancel a part of the factorizable one, more
precisely that proportional to $\frac{1}{N_c} \ c_1$ in $a_2$, leading to $a_2
\simeq c_2$.
Let us remind that such a cancellation seems to take place in two body  $D$
meson decays \cite{{R9},{R22}}.
However the issue is completely open in the $B$ meson sector.

\vs{2mm}
2$^{\bullet}$) \hs{3mm}
We now compute the four colour-suppressed two body decay rates
$$
B^o_d \ra \eta_c (J/\Psi) + K (K^*)
$$
assuming factorization and neglecting possible penguin contributions because at
least two or three gluons are needed in order to excite, from the vacuum, the
colourless charmonium state $c\ol{c}$.
The relevant colour-suppressed diagram is shown in Figure 1.

\vs{2mm}
\input /users/lpthe/keum/feynman.tex
\begin{center}
$$
\beginturtle
\shift{-190pt}{0pt}
\arrowtrue
\rotate{180}\putclabel{$\ol{b}$} \jump{-10pt}
\solidline{-90pt} \solidline{-90pt} \jump{-10pt} \putclabel{$\ol{c}$}
\shift{50pt}{-8pt} \putclabel{${\bf \eta_c} \hs{2mm}(J/\Psi)$}
\shift{-50pt}{8pt}
\jump{100pt} \rotate{120} \wavyline{60pt}{12}
\shift{-30pt}{20pt}
\putclabel{$W^+$}
\shift{30pt}{-20pt}
\shift{-180pt}{0pt} \putclabel{$B^o_d$}
\shift{180pt}{0pt}
\rotate{90} \solidline{70pt} \jump{10pt} \putclabel{$c$} \jump{-80pt}
\rotate{120} \solidline{-70pt} \jump{-10pt} \putclabel{$\ol{s}$}
\shift{0pt}{-15pt} \putclabel{$d$}
\rotate{-150}
\shift{50pt}{10pt} \putclabel{$K^o \hs{2mm}({K^o}^*)$} \shift{-50pt}{-10pt}
\jump{-10pt}  \solidline{-90pt}
\solidline{-90pt} \jump{-10pt} \putclabel{$d$}
\endturtle
$$

\vs{2mm}
Figure 1.
\end{center}

\vs{5mm}
The relevant Cabibbo-Kobayashi-Maskawa (CKM) factor is $V_{cb}^* V_{cs}$. The
hadronic form factors in the  BSW notation \cite{R9} are $F_0$ and $F_1$ for $B
\ra K$ transitions and $A_0, A_1, A_2$ and $V$ for $B \ra K^*$ transitions.
The decay matrix elements are proportional to the phenomenological
dimensionless parameter $a_2$ and involved the leptonic decay constants
$f_{\eta_c}$ and $f_{\Psi}$ defined as usual by :
\beqa
< \eta_c | \ol{c} \gamma_{\mu}\gamma_5 c| 0 > & = & i  \hs{1mm} f_{\eta_c}
\hs{1mm} p_{\mu} \label{eq5} \\
\cr
<J/\Psi | \ol{c} \gamma_{\mu} c | 0 > \hs{1mm} & = & f_{\Psi} \hs{1mm}
m_{\Psi} \hs{1mm} \epsilon_{\mu} \label{eq6}
\eeqa
where $p_{\mu}$ is the $\eta_c$ momentum and $\epsilon_{\mu}$ is the $J/\Psi$
polarization vector.

For the hadronic form factors it is convenient to consider separately their
normalizations at $q^2 = 0$ and their variations with $q^2$.

\vs{2mm}
\hs{5mm} $\ul{i)}$ for the normalization at $q^2 =0$,  we use the two well
known constraints \cite{R9}
\beqa
F_0(0) & = & F_1(0) \label{eq7} \\
\cr
A_0(0) & = & \frac{m_B + m_{K^*}}{2 m_{K^*}} \ A_1(0) - \frac{m_B - m_{K^*}}{2
m_{K^*}} \ A_2(0) \label{eq8}
\eeqa

Using $A_1^{BK^*}(0)$ as a scale, we define the three ratios
\beq
x = \frac{A_2^{BK^*}(0)}{A_1^{BK^*}(0)} \, \hs{10mm}
y = \frac{V^{BK^*}(0)}{A_1^{BK^*}(0)} \, \hs{10mm}
z = \frac{F_1^{BK}(0)}{A_1^{BK^*}(0)} \ \label{eq9}
\eeq
and from relations (7), (8) and (9), we get
\beqa
F_0^{BK}(0) & = & z \hs{2mm} {A_1^{BK^*}(0)} \label{eq10} \\
\cr
A_0^{BK^*}(0) & = & \frac{m_B + m_{K^*}}{2 m_{K^*}} \ [1 - \frac{m_B -
m_{K^*}}{m_B + m_{K^*}}\ \hs{1mm} x] \hs{2mm} {A_1^{BK^*}(0)} \label{eq11}
\eeqa

\vs{2mm}
\hs{5mm} $\ul{ii)}$ for the variation with $q^2$, we define the reduced
hadronic form factors normalized to unity at $q^2 = 0$
\beqa
f_j(q^2) & = & \frac{F_j(q^2)}{F_j(0)} \ \hs{15mm} {\rm for} \hs{2mm} j = 0, 1
\no \\
\cr
a_j(q^2) & = & \frac{A_j(q^2)}{A_j(0)} \ \hs{15mm} {\rm for} \hs{2mm} j = 0, 1,
2 \label{eq12} \\
\cr
v(q^2) & = & \frac{V(q^2)}{V(0)} \no
\eeqa
The type of variation with $q^2$ will be called in what follows a scenario.

\vs{3mm}
The four decay rates for $B^o_d \ra \eta_c (J/\Psi) + K (K^*)$ are conveniently
expressed in terms of the quantity $\Gamma_0$ given by
\beq
\Gamma_0 = [ \frac{G_F m_B^2}{\sqrt{2}} \ ]^2 \hs{2mm} |V_{cb}|^2 \hs{2mm}
|V_{cs}|^2 \hs{2mm} \frac{m_B}{16 \pi} \ a_2^2 \hs{2mm} |A_1^{BK^*}(0)|^2
\label{eq13}
\eeq
and we obtain the results
\beqa
& & \frac{\Gamma(B \ra \eta_c \hs{1mm} K)}{\Gamma_0} \  =
k_{\eta} \hs{1mm}
(1 - \frac{m_K^2}{m_B^2} \ )^2 \hs{1mm}
( \frac{f_{\eta_c}}{m_B} \ )^2 \hs{1mm}
|f_0(m^2_{\eta_c})|^2 \hs{1mm} z^2  \label{eq14} \\
\cr
& & \frac{\Gamma(B \ra J/\Psi \hs{1mm} K)}{\Gamma_0} \  =
(k_{\Psi})^3 \hs{1mm}
( \frac{f_{\Psi}}{m_B} \ )^2 \hs{1mm}
|f_1(m^2_{\Psi})|^2 \hs{1mm} z^2   \label{eq15} \\
\cr
& & \frac{\Gamma(B \ra \eta_c \hs{1mm} K^*)}{\Gamma_0} =
(k^*_{\eta})^3 \hs{1mm}
( \frac{m_B + m_{K^*}}{2 m_{K^*}} \ )^2 \hs{1mm}
( \frac{f_{\eta_c}}{m_B} \ )^2 \hs{1mm}
|a_0(m_{\eta_c}^2)|^2 \hs{1mm}
[1 - \frac{m_B - m_{K^*}}{m_B + m_{K^*}} \ \hs{1mm} x]^2  \label{eq16} \\
& & \frac{\Gamma(B \ra J/\Psi \hs{1mm} K^*)_{\ld\ld}}{\Gamma_0}  =
k^*_{\Psi} \hs{1mm}
( \frac{m_{\Psi}}{m_B} \ )^2 \hs{1mm}
(1 + \frac{m_{K^*}}{m_B} \ )^2 \hs{1mm}
( \frac{f_{\Psi}}{m_B} \ )^2 \hs{1mm}
| \hs{1mm} \{ \hs{3mm} \}_{\ld\ld} \hs{1mm} |^2  \label{eq17}
\eeqa
where for the longitudinal and transverse polarization of the final vector
mesons,  we have
\beqa
\{ \hs{3mm} \}_{LL} & = & a \hs{1mm} a_1(m^2_{\Psi}) - b \hs{1mm}
a_2(m^2_{\Psi}) \hs{1mm} x \label{eq18} \\
\cr
\{ \hs{3mm} \}_{\pm\pm} & = &  a_1(m^2_{\Psi}) \mp \hs{1mm} c
\hs{1mm} v(m^2_{\Psi}) \hs{1mm} y \label{eq19}
\eeqa

The dimensionless coefficients $a, b$, and $c$ depend only on the masses
\cite{R1} :
\beqa
a & = & \frac{m_B^2 - m_{\Psi}^2 - m_{K^*}^2}{2 \hs{1mm} m_{\Psi} \hs{1mm}
m_{K^*} } \ \label{eq20} \\
\cr
b & = & \frac{(k^*_{\Psi})^2 \hs{1mm} m_B^4}{2 \hs{1mm} m_{\Psi} \hs{1mm}
m_{K^*} \hs{1mm} (m_B + m_{K^*})^2 } \ \label{eq21} \\
\cr
c & = & \frac{k^*_{\Psi} \hs{1mm} m_B^2}{ (m_B + m_{K^*})^2 } \ \label{eq22}
\eeqa
where the dimensionless parameters \hs{1mm} $k$ \hs{1mm} in Eqs.(14) - (22) are
simply related to the - center of mass - momentum $K$ of the final mesons by $k
= 2 K/m_B$.

Inserting now the masses values, we get for $B^o$ decay \footnote{ The
numerical coefficients in Eqs.(23)-(34) are slightly different  for $B^+$ decay
because of the $K^o - K^+$ and ${K^*}^o - {K^*}^+$ mass differences. We shall
ignore this effect in this paper.}
\beqa
& & \frac{\Gamma(B \ra \eta_c \hs{1mm} K)}{\Gamma_0}  =  0.652 \hs{1mm}
( \frac{f_{\eta_c}}{m_B} \ )^2 \hs{1mm}
|f_0(m^2_{\eta_c})|^2 \hs{1mm} z^2  \label{eq23} \\
\cr
& & \frac{\Gamma(B \ra J/\Psi \hs{1mm} K)}{\Gamma_0}  =  0.259 \hs{1mm}
( \frac{f_{\Psi}}{m_B} \ )^2 \hs{1mm}
|f_1(m^2_{\Psi})|^2 \hs{1mm} z^2   \label{eq24} \\
\cr
& & \frac{\Gamma(B \ra \eta_c \hs{1mm} K^*)}{\Gamma_0} =  2.884 \hs{1mm}
( \frac{f_{\eta_c}}{m_B} \ )^2 \hs{1mm}
|a_0(m_{\eta_c}^2)|^2 \hs{1mm}
[1 - 0.711 \hs{1mm} x]^2  \label{eq25} \\
\cr
 & & \frac{\Gamma(B \ra J/\Psi \hs{1mm} K^*)_{\ld\ld}}{\Gamma_0}  =
0.280 \hs{1mm}
( \frac{f_{\Psi}}{m_B} \ )^2 \hs{1mm}
| \hs{1mm} \{ \hs{3mm} \}_{\ld\ld} \hs{1mm} |^2  \label{eq26}
\eeqa
with
\beqa
\{ \hs{3mm} \}_{LL} & = & 3.148 \hs{1mm} a_1(m^2_{\Psi}) \hs{1mm}
[1 - 0.412 \hs{1mm} \frac{a_2(m^2_{\Psi})}{a_1(m^2_{\Psi})} \ \hs{1mm} x]
\label{eq27} \\
\cr
\{ \hs{3mm} \}_{\pm\pm} & = &  a_1(m^2_{\Psi}) \hs{1mm}
[1 \mp \hs{1mm} 0.434 \hs{1mm}
\frac{v(m^2_{\Psi})}{a_1(m^2_{\Psi})} \ \hs{1mm} y] \label{eq28}
\eeqa
In particular :
\beq
\frac{\Gamma(B \ra J/\Psi K^*)_{LL}} {\Gamma_0} =
2.773 \hs{1mm} ( \frac{f_{\Psi}}{m_B} \ )^2 \hs{1mm}
|a_1(m^2_{\Psi})|^2 \hs{1mm}
[1 - 0.412 \hs{1mm} x(m^2_{\Psi}) ]^2 \label{eq29}
\eeq
where, as used in \cite{R1}
\beq
x(m^2_{\Psi}) = \frac{a_2(m^2_{\Psi})}{a_1(m^2_{\Psi})} \
\hs{1mm} x . \label{eq30}
\eeq

For the comparison of $\eta_c$ and $\Psi$ production, we introduce the two
ratios of rates
\beqa
R_K \equiv
\frac{\Gamma(B \ra \eta_c \hs{1mm} K)}{\Gamma(B \ra J/\Psi \hs{1mm} K)} \
&=&
2.520 \hs{1mm}
( \frac{f_{\eta_c}}{f_{\psi}} \ )^2 \hs{1mm}
[ \frac{f_0(m^2_{\eta_c})}{f_1(m^2_{\Psi})} \ ]^2 \label{eq31} \\
\cr
R_{K^*} \equiv
\frac{\Gamma(B \ra \eta_c \hs{1mm} K^*)}{\Gamma(B \ra J/\Psi \hs{1mm} K^*)} \
&=&
1.040 \hs{1mm}
( \frac{f_{\eta_c}}{f_{\psi}} \ )^2 \hs{1mm}
[ \frac{a_0(m^2_{\eta_c})}{a_1(m^2_{\Psi})} \ ]^2 \hs{1mm}
\frac{(1 - 0.711 \hs{1mm} x )^2 \hs{2mm} \rho_L}
{[1 - 0.412 \hs{2mm}x(m^2_{\Psi})]^2} \ \label{eq32}
\eeqa
where $\rho_L$ has been defined in Eq.(2).

The two ratios $R_K$ and $R_{K^*}$ suffer of our ignorance of the scalar decay
constant $f_{\eta_c}$ and such  uncertainty can be eliminated by considering
the relative rate of $K^*$ versus $K$ production.
\beq
R_{\eta_c} \equiv
\frac{\Gamma(B \ra \eta_c \hs{1mm} K^*)}{\Gamma(B \ra \eta_c \hs{1mm} K)} \ =
\frac{R_{K^*}}{R_K} \ \hs{1mm} R_{\Psi} \label{eq33}
\eeq

Using Eqs.(31), (32) and (33), we obtain
\beq
R_{\eta_c} = 0.413 \hs{1mm}
[ \frac{a_0(m^2_{\eta_c})}{a_1(m^2_{\Psi})} \ ]^2 \hs{1mm}
[ \frac{f_1(m^2_{\Psi})}{f_0(m^2_{\eta_c})} \ ]^2 \hs{1mm}
\frac{(1 - 0.711 \hs{1mm} x )^2 \hs{2mm} \rho_L}{[1 - 0.412 x(m^2_{\Psi})]^2} \
\hs{2mm} R_{\Psi} \label{eq34}
\eeq

This measurable quantity $R_{\eta_c}$ is to be used to test the factorization
hypothesis.

We observe that the normalization parameters $y$ and $z$ do not appear in the
ratios $R_K$, $R_{K^*}$ and $R_{\eta_c}$, only $x$ being involved in $R_{K^*}$
and $R_{\eta_c}$.
The physical meaning of this fact can be understood in the following way :
by angular momentum conservation, the $K^*$ is longitudinally polarized in $B
\ra K^* + \eta_c$, hence there exists a link between this rate and the
longitudinal one $\Gamma(B \ra K^* J/\Psi)_{LL}$, the latter depends only on
$x$ (not on $y$) as can be seen in Eq.(29).

\vs{5mm}
\hspace{3mm} \large{} {\bf III. \hs{3mm} Bounds for the ratio ${\bf R_{\eta_c}}
$}    \vspace{0.5cm}
\normalsize

$1^{\bullet}$) \hs{2mm} As shown in \cite{R1}, for a fixed value of $\rho_L$,
the quantity $x(m^2_{\Psi})$ is restricted to lie in the range limited by an
arc of hyperbola :
\beq
0 \hs{2mm} \leq \hs{2mm} x(m^2_{\Psi}) \hs{2mm} \leq \hs{2mm} x_{max}(\rho_L)
\eeq
with
\beq
x_{max}(\rho_L) = \frac{1}{b} \ [ \hs{2mm}a - \sqrt{ \frac{2 \rho_L}{1 -
\rho_L}} \hs{2mm} ]
\eeq
where the numerical values of the coefficients $a$
and $b$ have been given in Eq.(27) : \\
$a = 3.148, \hs{2mm} b/a = 0.412$.

The positivity of $x_{max}(\rho_L)$ implies a theoretical upper bound
$\rho^{max}_L$ for $\rho_L$ given by \cite{R1} $ : \\
\rho^{max}_L = a^2/(a^2 + 2) $, its numerical value is  $\rho^{max}_L = 0.832$.

In particular from Eq.(36), we get
\beq
\frac{\rho_L}{[1 - \frac{b}{a} \ \hs{1mm} x_{max}(\rho_L)]^2} \  \hs{2mm} =
\hs{2mm} \frac{a^2}{2} \ (1 - \rho_L) \label{eq36}
\eeq
This result is scenario-independent.

\vs{2mm}
$2^{\bullet}$) \hs{2mm} However the relation between $x(m^2_{\Psi})$ and
\hs{1mm} $x$ \hs{1mm} is scenario-dependent and there are three possibilities :
$$
{\rm (i)} \hs{2mm} x(m^2_{\Psi}) \hs{1mm} = \hs{1mm} x \hs{4mm}, \hs{4mm}
{\rm (ii)} \hs{2mm} x(m^2_{\Psi}) \hs{1mm} > \hs{1mm} x \hs{4mm}, \hs{4mm}
{\rm (iii)} \hs{2mm} x(m^2_{\Psi}) \hs{1mm} < \hs{1mm} x .
$$
Among the three possibilites, only the first two cases seem to be physically
plausible.
The case (i) occurs when both axial form factors $A_1^{BK^*}(q^2)$ and
$A_2^{BK^*}(q^2)$ have the same $q^2$ dependence no matter how it is :
monopole, dipole, exponential form, etc.
Many commonly used models sit in this scenarios
\cite{{R9},{R11},{R12},{R14},{R15},{R18}}.
Of course constant axial form factors belong to this case.

In the case (ii), the ratio $A_2^{BK^*}(q^2)/A_1^{BK^*}(q^2)$ is an increasing
function of $q^2$ and such a situation is realized when $A_1^{BK^*}(q^2)$
varies more slowly than $A_2^{BK^*}(q^2)$ \cite{{R8},{R10},{R16},{R17}}, for
instance, when $A_1$ decreases with $q^2$ and $A_2$ is either constant or
increasing with $q^2$, or when $A_1$ is constant and $A_2$ is increasing.

The cases (i) and (ii) obviously come from quark models, $QCD$ sum rules, and
lattice calculations.
To our knowledge, the case (iii) has never been met by analyses in the
literature (see an extensive discussion in Ref.\cite{R8}) and it will be
disregarded in what follows where we shall use $x(m^2_{\Psi}) \geq  x$.

We emphasize here that we do not have to assume anything about the individual
form factors $A_1(q^2)$, $A_2(q^2)$, neither their normalizations at $q^2 = 0$,
nor their long extrapolation from $q^2 = 0$ to $q^2 = m^2_{\Psi}$, both being
sources of uncertainties.
Only the ratio $a_2(m^2_{\Psi})/a_1(m^2_{\Psi})$ at one fixed value $q^2 =
m_{\Psi}^2$ enters in  our discussion.

\vs{2mm}
$3^{\bullet}$) \hs{2mm} The function $(1 - 0.711 \hs{1mm} x)^2$ is a non
negative function of $x$ which decreases for $x < 1.41$, vanishes at $x = 1.41$
and increases for $x > 1.41$.
For $x(m^2_{\Psi}) \geq x $,  we get
\beq
[1 - 0.711 \hs{1mm} x_{max}(\rho_L)]^2 \hs{2mm} \leq \hs{2mm} (1 - 0.711
\hs{1mm} x)^2
\hs{2mm} \leq \hs{2mm} 1
\label{eq37}
\eeq
The lower bound vanishes at $\rho_L = 0.468$ and at this point $x_{max}(\rho_L)
= 1.41$.

\vs{2mm}
$4^{\bullet}$) \hs{2mm} The $x$ dependence of the ratios $R_{K^*}$ and
$R_{\eta_c}$ in Eqs.(32) and (34) is given by  the quantity

\beq
P(\rho_L) = \frac{(1 - 0.711 \hs{1mm} x)^2 \hs{2mm} \rho_L}{[1 - 0.412 \hs{1mm}
x(m^2_{\Psi})]^2} \ \label{eq38}
\eeq

\vs{2mm}
For $x(m^2_{\Psi}) \geq x $, at a fixed value of $\rho_L \leq \rho^{max}_L$,
the upper bound of $P(\rho_L)$ corresponds to $x(m^2_{\Psi}) = x = 0$
and the lower bound of $P(\rho_L)$ to $ x(m^2_{\Psi}) = x = x_{max}(\rho_L)$.

Using now Eqs.(37) and (38), we get

\beq
[1 - 0.711 \hs{1mm} x_{max}(\rho_L)]^2 \hs{1mm} \frac{a^2}{2} \ \hs{1mm}(1 -
\rho_L)
\hs{2mm} \leq \hs{2mm} P(\rho_L) \hs{2mm} \leq \hs{2mm} \rho_L \label{eq39}
\eeq
Of course the lower and upper bounds of $P(\rho_L)$ coincide when $\rho_L$
takes its maximal theoretical value $\rho^{max}_L = 0.832$.
We present in Figure 2 the bounds of the quantity $P(\rho_L)$ as a function of
$\rho_L$ for $\rho_L^{max} \geq \rho_L \geq 0.4 $.

\vs{2mm}
$5^{\bullet}$) \hs{2mm} In the expressions of $R_{K}$ and $R_{K^*}$ we have two
scenario-dependent factors
\beq
S_V = [ \frac{f_0(m^2_{\eta_c})}{f_1(m^2_{\Psi})} \ ]^2 \hs{20mm}
S_A = [ \frac{a_0(m^2_{\eta_c})}{a_1(m^2_{\Psi})} \ ]^2
\eeq

Two preliminary remarks are in order. Firstly we are considering ratios of form
factors $S_V(q^2), S_A(q^2)$ and their dependences on $q^2$ are expected to be
smooth. Secondly the $\eta_c$ and $J/\Psi$ masses being close to each other,
the two ratios $S_V$ and $S_A$ are much less uncertain than the individual
$f_0, f_1, a_0, a_1$ extrapolated lengthly from $q^2 = 0$ to $m_{\eta_c^2}$ or
$m^2_{\Psi}$.
Therefore the quantities $S_V$ and $S_A$ are expected to be $O(1)$.

In order to have a more precise idea about $S_V$ and $S_A$, let us compute
these quantities using a monopole $q^2$ dependence for all form factors with
the values of the pole masses as proposed by BSW \cite{R9}, $m_{1^-} = 5.43
\hs{1mm} GeV$, $m_{0^+} = 5.89 \hs{1mm} GeV $, $m_{1^+} = 5.82 \hs{1mm} GeV$
and $m_{0^-} = 5.38 \hs{1mm} GeV$.
The result is
\beq
S_V = 0.822 \hs{20mm} S_A = 1.069 \label{eq41}
\eeq
and for the quantity entering in $R_{\eta_c}$
\beq
\frac{S_A}{S_V} \ = 1.300 \label{eq42}
\eeq
The same calculation has been repeated with a large number of scenarios
\cite{R9}$-$\cite{R18} and it turns out that the results are only weakly
scenario dependent.
In what follows we shall use a conservative range
\beq
 1 \hs{2mm} \leq \hs{2mm} \frac{S_A}{S_V} \ \hs{2mm} \leq \hs{2mm} 1.4
\label{eq43}
\eeq

\vs{2mm}
$6^{\bullet}$) \hs{2mm} The relation between the ratios of rates $R_{\eta_c}$
and $R_{\Psi}$ is a function of $\rho_L$ given by
\beq
R_{\eta_c} = 0.413 \hs{1mm} \frac{S_A}{S_V} \ \hs{1mm} P(\rho_L)
\hs{1mm} R_{\Psi} \label{eq44}
\eeq
We plot in Figure 3 the quantity $R_{\eta_c}/R_{\Psi}$ as a function of
$\rho_L$ for $\rho^{max}_L \geq \rho \geq 0.4 $ using the theoretical estimate
of $S_A/S_V$ given in Eq.(44), i.e. $S_A/S_V = 1$ for the lower bound and
$S_A/S_V = 1.4$ for the upper bound.
We also indicate the one standard deviation lower value of $\rho_L$ coming from
CLEO II \cite{R3}, i.e. $\rho_L \geq 0.71$ and the CDF preliminary result
within one standard deviation $0.53 < \rho_L < 0.80$ \cite{R5}.

In order to obtain numerical estimates, let us choose $\rho_L = 0.71 $.
{}From Figure 3, we get
\beq
0.14 \leq \frac{R_{\eta_c}}{R_{\Psi}} \ \leq 0.41 \label{eq45}
\eeq
If $\rho_L$ increases, both lower and upper bounds in Eq.(45) increase, and for
$\rho_L = \rho^{max}_L$ we get from Figure 3 :
\beq
0.34 \leq \frac{R_{\eta_c}}{R_{\Psi}} \ \leq 0.48 \label{eq46}
\eeq
Combining two Eqs.(46), (47) and using CLEO data on $\rho_L$ ($0.71 \leq \rho_L
\leq \rho^{max}_L $), we obtain
\beq
0.14 \leq \frac{R_{\eta_c}}{R_{\Psi}} \ \leq 0.48 \label{eq47}
\eeq
Bounds on the ratio $R_{\eta_c}$ are now obtained using the experimental value
of $R_{\Psi}$, with $R_{\Psi} = 1.64 \pm 0.34$ \cite{R2} we get
\beq
0.18 \leq R_{\eta_c} \leq 0.95 . \label{eq48}
\eeq
Using $R_{\Psi} = 1.71 \pm 0.34$ \cite{R3}, we obtain similar results :
\beq
0.19 \hs{2mm} \leq \hs{2mm} R_{\eta_c} \hs{2mm} \leq \hs{2mm} 0.98 \label{eq49}
\eeq
Our form factor model-independent result Eq.(48) could be favourably compared
to
the model-dependent one recently discussed in Ref.\cite{R7}.
Using their Eqs.(17), (25) and their Table 1, the three models of form factor
considered there \cite{R7} (which correspond respectively to our
Refs.\cite{{R9},{R18},{R15}}) yield respectively for $R_{\eta_c}/R_{\Psi} $ the
following values : $0.071, 0.047$ and $0.38$.
The first two models violate our bounds Eq.(48), the third one \cite{R15}
(constructed to fit large $\rho_L$ data) fails badly to account for $R_{\Psi}$.
Note also that the first two models fail to explain either $R_{\Psi}$ or
$\rho_L$ data \cite{R1}.
Using these models to compute the $B \ra \eta_c + K^*$ rate is therefore
questionable.

\vs{5mm}
\hs{3mm} \large{ } {\bf IV. \hs{3mm} Determination of ${\bf f_{\eta_c}}$ }
\normalsize
\vs{3mm}

$1^{\bullet}$) \hs{2mm}
A measurement of the ratio $R_K$ will provide an opportunity to extract, from
experiment, the scalar decay constant $f_{\eta_c}$.
Assuming factorization, from Eq.(31) we obtain
\beq
f_{\eta_c} = 0.63 \hs{2mm} \sqrt{ \frac{R_K}{S_V} \ } \hs{2mm} f_{\Psi}
\label{eq53}
\eeq
The vector constant $f_{\Psi}$ has been estimated from the decay rate $J/\Psi
\ra e^+ e^-$ to be $f_{\Psi} = 382 \hs{1mm} MeV$ \cite{R10}.
The scenario-dependent factor $S_V$ discussed
in the previous section is choosen as
\beq
S_V = 0.8 \pm 0.1 \label{eq54}
\eeq
and we get
\beq
f_{\eta_c} = \sqrt{R_K} \hs{2mm} [254 - 288] \hs{2mm} MeV \label{eq55}
\eeq

\vs{2mm}
$2^{\bullet}$) \hs{2mm}
We have theoretical ways to estimate both $f_{\Psi}$ and $f_{\eta_c}$ constants
using the wave function of the charmonium $S$ state at the origin \cite{R7}
\beq
f^2_{\Psi} = \frac{12}{m_{\Psi}} \ |\Psi(0)|^2 \hs{20mm}
f^2_{\eta_c} =  48 \hs{1mm} \frac{m_c^2}{m^3_{\eta_c}} \ |\Psi(0)|^2
\label{eq56}
\eeq
and the ratio of the two decay constants takes the form
\beq
( \frac{f_{\eta_c}}{f_{\Psi}} \ )^2 = 4 \hs{1mm} \frac{m^2_c \hs{2mm}
m_{\Psi}}{m^3_{\eta_c}} \ \label{eq57}
\eeq
Using $m_c =1.45 \hs{1mm} GeV$ and the experimental value for $m_{\eta_c}$ and
$m_{\Psi}$, we obtain :
\beq
\frac{f_{\eta_c}}{f_{\Psi}} \simeq  0.993  \label{eq58}
\eeq
$QCD$ corrections can modify both estimates (54) for $f_{\eta_c}$ and
$f_{\Psi}$, but they are expected to approximately cancel in the ratio
$f_{\eta_c}/f_{\Psi}$ \cite{R23}.
Using the value of Eq.(56) and the previous estimate of the scenario-dependent
factor $S_V$, a prediction for the ratio $R_K$ due to factorization can be
obtained :
\beq
R_K = 1.8 \hs{2mm} - \hs{2mm} 2.3 \label{eq59}
\eeq

\vs{2mm}
$3^{\bullet}$) \hs{2mm}
Analogous considerations can be made for the ratio $R_{K^*}$ which from Eq.(32)
is written as
\beq
R_{K^*} = 1.040 \hs{2mm} ( \frac{f_{\eta_c}}{f_{\Psi}} \ )^2 S_A \hs{2mm}
P(\rho_L)
\label{eq60}
\eeq
However we have a supplementary uncertainty due to the polarization factor
$P(\rho_L)$
and the less accurately known scenario-dependent factor $S_A$.
Using $S_A = 1 \pm 0.2 $ and the results previously obtained in Figure 2 for
$P(\rho_L)$, the prediction for $R_{K^*}$, similar to Eq.(57) for $R_K$, is
\beq
R_{K^*} = 0.28 \hs{2mm} - \hs{2mm} 1.05 \label{eq61}
\eeq
where, using the CLEO II data, we have taken $\rho_L \geq 0.71$.

\vs{2mm}
$4^{\bullet}$) \hs{2mm}
However the best place to look for the determination of $f_{\eta_c}$ is
probably the inclusive $B$ meson decay $B \ra \eta_c + X_s$ which is presumably
described at the quark level by $b \ra \eta_c + s$ as thoroughly discussed in
Ref.\cite{R7} where the following expression has been obtained \cite{R7} :
\beq
\frac{ \Gamma(B \ra \eta_c X_s)}{\Gamma(B \ra J/\Psi X_s)} \ =
( \frac{f_{\eta_c}}{f_{\Psi}} \ )^2 \frac{\tilde{K}_{\eta_c}}{\tilde{K}_{\Psi}}
\
\frac{[(m_b^2 - m_s^2)^2 - (m_b^2 + m_s^2) \hs{1mm} m_{\eta_c}^2]}
{[(m_b^2 - m_s^2)^2 \hs{1mm} + \hs{1mm} m_{\Psi}^2 \hs{1mm} (m_b^2 + m_s^2 - 2
\hs{1mm} m_{\Psi}^2)]} \ \label{eq62}
\eeq
 $\tilde{K}_{\eta_c}$ and $\tilde{K}_{\Psi}$ are the $\eta_c$ and $J/\Psi$ -
center of mass - momenta  corresponding respectively to the decays $b \ra
\eta_c + s$ and $b \ra J/\Psi + s$.

The expression of Eq.(60) is independent on hadronic form factors.
However it involves the two quark masses $m_b, m_s$ and  is therefore sensitive
to their numerical values.
In particular the value of $m_s$ is not accurately known and it could be a
source of uncertainty.
Using $m_b =   4.8 \hs{1mm} GeV$, and $ m_s = 0.5 \hs{1mm} GeV$, we obtain
\beq
\frac{ \Gamma(B \ra \eta_c X_s)}{ \Gamma(B \ra J/\Psi X_s)} \ = 0.59 \hs{2mm} (
\frac{f_{\eta_c}}{f_{\Psi}} )^2 \label{eq63}
\eeq

\vs{5mm}
\hspace{3mm} \large{} {\bf V. \hs{3mm} Summary and Conclusion}
\vspace{3mm}
\normalsize

The large discrepancy recently found \cite{R1} between $B \ra K(K^*) + J/\Psi$
experiments and theoretical predictions - using the factorization hypothesis
together with form factors models - prompts us to ask the question : is this
due to the materials (form factors) or to the methodology (factorization) ?
Assuming only the latter and not the formers, we can predict the $B \ra K(K^*)
+ \eta_c$ rates with the help of $B \ra K(K^*) + J/\Psi$ data.
 Experimental measurement of the decay rates for $\eta_c$ production provides
then a genuine test of the factorization assumption.
We point out that such a test is free of uncertainties (due to both final state
inelastic interactions and $W$-exchange diagram contributions) that usually
contaminate the analogous test in the charm sector.
The reason for the absence of final state inelastic interactions is that
in $B \ra K(K^*) + J/\Psi$, there is only one  isospin 1/2 decay amplitude,
while for Cabibbo favoured charm decay, both isospin 1/2 and 3/2 amplitudes
interfer in the relatively low charm mass region.

The ratio $R_{\eta_c} \equiv \frac{B \ra K^* + \eta_c}{B \ra K + \eta_c} \ $,
predicted by factorization, is unexpectedly low between 0.18 and 0.95 using the
CLEO II results \cite{R3} and even lower if the CDF longitudinal fraction
$\rho_L$ \cite{R5} is prefered.
Its confirmation or not by experiment will fix up the status of factorization
in colour-suppressed beauty decays.

The problem of the experimental determination of the scalar constant
$f_{\eta_c}$ has been discussed together with theoretical expectation.
Again experiment will clarify the situation if the test of factorization for
$R_{\eta_c}$ is successfully passed.

In the appendix we also discuss the question of the transverse polarization of
the final vector mesons in the decay $B \ra J/\Psi + K^*$.

\vspace{5cm}
\hspace{1cm} \Large{} {\bf Acknowledgements}    \vspace{0.5cm}
\normalsize

\vs{2mm}
Y. Y. K would like to thank the Commissariat \`a l'Energie Atomique of France
for the award of a fellowship
and especially G. Cohen-Tannoudji and J. Ha\"{\i}ssinski
for their encouragements.

\newpage
\hs{5mm} \Large{ } {\bf APPENDIX}
\vs{5mm}

\large{ } {\bf Comment on the transverse polarization in $B \ra J/\Psi + K^*$ }
\normalsize \vs{3mm}

The bounds shown in Fig. 3 will be  improved if a full measurement of the
polarizations in the processes $B \ra J/\Psi + K^*$ can be experimentally
achieved.
For that purpose the two transverse polarizations have to be separated and the
left-right helicity asymmetry  ${\cal A}_{LR}$ defined by

\beq
\cA_{LR} = \frac{ \Gamma(B \ra J/\Psi + K^*)_{--} - \Gamma(B \ra J/\Psi +
K^*)_{++} }
{ \Gamma(B \ra J/\Psi + K^*)_{--} + \Gamma(B \ra J/\Psi + K^*)_{++} } \
\label{A1}
\eeq
determines the quantity $y(m^2_{\Psi})$ \cite{R1} related to \hs{1mm} $y$
\hs{1mm} by
\beq
y(m^2_{\Psi}) = \frac{v(m^2_{\Psi})}{a_1(m^2_{\Psi})} \ \hs{2mm} y
\label{A2}
\eeq
Using Eq.(19) we obtain
\beq
\cA_{LR} = \frac{ 2 \hs{1mm} c \hs{1mm} y(m^2_{\Psi}) }{1 + c^2 \hs{1mm}
y^2(m^2_{\Psi}) } \label{A3}
\eeq
and the bounds on $\cA_{LR}$ for $y(m^2_{\Psi}) \geq 0$
- we notice that $y \geq 0$ is characteristic of the $V - A$ current
while for a $V + A$ current $y$ is negative -
 are simply
\beq
0 \hs{2mm} \leq \hs{2mm} \cA_{LR} \hs{2mm} \leq \hs{2mm} 1. \label{A4}
\eeq
The measurement of $\cA_{LR}$ determines for $y(m^2_{\Psi})$ two solutions :
\beq
c\hs{1mm} y(m^2_{\Psi}) = \frac{ 1 \pm \sqrt{1 - \cA_{LR}^2}} {\cA_{LR}} \
\label{A4}
\eeq
The numerical value of \hs{1mm} $c$ \hs{1mm} as given by Eq.(28) is 0.434.

On the other hand, as discussed in \cite{R1}, for a given value of $\rho_L$,
the allowed domain of $x(m^2_{\Psi})$ and $y(m^2_{\Psi})$ are located on an arc
of hyperbola whose limiting points are
\beqa
 y(m^2_{\Psi}) = 0 \hs{5mm}, & & \hs{4mm} x(m^2_{\Psi}) = x_{max}(\rho_L) \no
\\
\cr
 x(m^2_{\Psi}) = 0 \hs{5mm}, & & \hs{4mm} y(m^2_{\Psi}) = y_{max}(\rho_L) \no
\eeqa
The value of $x_{max}(\rho_L)$ has been given in Eq.(36) and for
$y_{max}(\rho_L)$ we obtain :
\beq
y_{max}(\rho_L) = \frac{1}{c} \ \sqrt{ \frac{a^2 + 2}{2} \ } \sqrt{
\frac{\rho^{max}_L}{\rho_L} \ - 1}
= 5.623 \hs{2mm} \sqrt{ \frac{\rho^{max}_L}{\rho_L} \ - 1} \label{A6}
\eeq
The measurements of the longitudinal and transverse polarizations in $B \ra
J/\Psi + K^*$ are compatible if and only if the value of $y(m^2_{J/\psi})$
obtained from a measurement
of the asymmetry $\cA_{LR}$ satisfies the inequality $y(m^2_{J/\psi}) \leq
y_{max}(\rho_L)$.

It is straightforward to check that when
\beq
0.712 = \frac{a^2}{a^2 + 4} \ \hs{2mm} \leq \hs{2mm} \rho_L \hs{2mm} \leq
\hs{2mm} \rho^{max}_L = \frac{a^2}{a^2 + 2} \ = 0.832 \label{A7}
\eeq
the upper bound of $\cA_{LR}$ is not 1 but a quantity $\cA_{LR}^{max} \leq 1$
given by :

\beq
\cA_{LR}^{max} = \frac{2}{a^2} \ \sqrt{2 \hs{1mm} (a^2 + 2)} \frac{
\sqrt{\rho_L(\rho_L^{max} - \rho_L)}} {(1 - \rho_L)} \
= 0.985 \hs{2mm} \frac{ \sqrt{\rho_L(\rho_L^{max} - \rho_L)}}{(1 - \rho_L)} \
 \label{A8}.
\eeq
{}From experiment this range of  $\rho_L$ between 0.712 and 0.832 is a
realistic one.

%
%
%
%

\newpage
%

\newpage
\section*{Figure captions}
\normalsize
\vspace{0.5cm}

\ben

\item
{\bf Fig. 1} : \hs{3mm}
The colour-suppressed Feynman diagram of $B^o_d \ra \eta_c(J/\Psi) + K(K^*)$
decay processes.
 \\

\item
{\bf Fig. 2} : \hs{3mm}
Bounds for the polarization dependent function $P(\rho_L)$ as functions of
$\rho_L$.
 \\
\item
{\bf Fig. 3} : \hs{3mm}
Bounds for the ratio $R_{\eta_c}/R_{\Psi}$ as functions of $\rho_L$.
The full vertical line is the CLEO II lower bound on $\rho_L$.
The dotted vertical lines are the one standard deviation $\rho_L$ of CDF.
 \\

\een

\end{document}